\documentclass[12pt]{article}%
\usepackage{amsmath}
\usepackage{amsfonts}
\usepackage{amssymb}
\usepackage{graphicx}
\begin{document}
\begin{titlepage}
\begin{center}
\renewcommand{\thefootnote}{\fnsymbol{footnote}}
{\Large\bf Consistent Massive Colored Gravitons} \vskip20mm
{\large\bf{Ali H.
Chamseddine \footnote{email: chams@aub.edu.lb}}}\\
\renewcommand{\thefootnote}{\arabic{footnote}}
\vskip2cm {\it Center for Advanced Mathematical Sciences (CAMS)
and\\
Physics Department, American University of Beirut, Lebanon.\\}
\end{center}
\vfill
\begin{center}
{\bf Abstract}
\end{center}
\vskip 1cm A short review of the problems with the action for
massive gravitons is presented. We show that consistency problems
could be resolved by employing spontaneous symmetry breaking to
give masses to gravitons. The idea is then generalized by
enlarging the $SL(2,\mathbb{C})$ symmetry to
$SL(2N,\mathbb{C})\times SL(2N,\mathbb{C})$ which is broken to
$SL(2,\mathbb{C})$ spontaneously through a non-linear realization.
The requirement that the space-time metric is generated
dynamically forces the action constructed to be a four-form. It is
shown that the spectrum of this model consists of  two sets of
massive matrix gravitons in the adjoint representation of $SU(N)$
and thus are colored, as well as two singlets, one describing a
massive graviton, the other being the familiar massless graviton.
\end{titlepage}

\section{Introduction}

Massive gravitons occur in higher dimensional theories of gravity when
compactified to lower dimensions as an infinite tower whose masses are
multiples of the Planck mass \cite{scherk}. They are referred to as
Kaluza-Klein (KK) gravitons. They also occur in brane models where it is
possible to have an effective four-dimensional theory with ultra-light
gravitons in addition to a massless graviton \cite{DamourKogan}. These are
referred to as multigravity models. They are also present in noncommutative
geometry for spaces which are products of a discrete space of two or more
points times a manifold \cite{Connes}. This can be realized as a multi-sheeted
space, e.g. when $X=Y\times Z_{2}$, there will be a metric on every sheet,
resulting in a bigravity model \cite{CFF}

In the KK approach the higher dimensional metric is expanded in terms of the
compactified coordinates. As an example, in five dimensions%
\[
g_{\mu\nu}\left(  x,y\right)  =%
%TCIMACRO{\dsum \limits_{n=0}^{\infty}}%
%BeginExpansion
{\displaystyle\sum\limits_{n=0}^{\infty}}
%EndExpansion
g_{\mu\nu n}\left(  x\right)  e^{iny},
\]
where $y$ \ is the compact coordinate in the fifth dimension. Besides the zero
mode (massless graviton), there will be massive modes (massive gravitons) of
masses $nM_{P}$ where $M_{P}$ is the Planck mass. It is not possible to obtain
a graviton with a very small mass in the KK approach. This, however, is
possible in brane models \cite{RS} \cite{DamourKogan} and noncommutative
geometric models \cite{CFF} \cite{Lizzi}.

A Lagrangian for massive spin-2 field was proposed long ago by Fierz and Pauli
\cite{FP}. It is given by
\begin{align*}
I  &  =-\frac{1}{4}%
%TCIMACRO{\dint }%
%BeginExpansion
{\displaystyle\int}
%EndExpansion
d^{4}x\left(  \partial_{\lambda}h_{\mu\nu}\partial^{\lambda}h^{\mu\nu
}-2\partial^{\nu}h_{\mu\nu}\partial_{\lambda}h^{\mu\lambda}+2\partial^{\nu
}h_{\mu\nu}\partial^{\mu}h_{\lambda}^{\;\lambda}-\partial_{\mu}h_{\nu
}^{\,\;\nu}\partial^{\mu}h_{\lambda}^{\;\lambda}\right. \\
&  \hspace{0.7in}\left.  +m^{2}\left(  h_{\mu\nu}h^{\mu\nu}-bh_{\nu}^{\,\;\nu
}h_{\lambda}^{\;\lambda}\right)  \right)  ,
\end{align*}
where, for consistency, $b=1$ which guarantees that only five components of
$h_{\mu\nu}$ propagate, instead of the expected six. The $m$ independent part
is the same as that obtained by linearizing the Einstein-Hilbert action, which
is the Lagrangian for a massless spin-2 field in a Minkowski background. The
propagator for the massive graviton $h_{\mu\nu}$ is \cite{vandam} \cite{Deser}
\cite{ISS1},
\begin{align*}
\Delta_{\mu\nu}^{\rho\sigma}  &  =\frac{1}{m^{2}-k^{2}}\left(  \left(
\delta_{\mu}^{\rho}-\frac{k_{\mu}k^{\rho}}{m^{2}}\right)  \left(  \delta_{\nu
}^{\sigma}-\frac{k_{\nu}k^{\sigma}}{m^{2}}\right)  \right. \\
&  \hspace{0.5in}-\frac{1}{3}\left(  \eta_{\mu\nu}-\frac{k_{\mu}k_{\nu}}%
{m^{2}}\right)  \left(  \eta^{\rho\sigma}-\frac{k^{\rho}k^{\sigma}}{m^{2}%
}\right) \\
&  \hspace{0.5in}\left.  +\frac{1-b}{2\left(  1-b\right)  k^{2}+\left(
1-4b\right)  m^{2}}\left(  \eta_{\mu\nu}+\frac{2k_{\mu}k_{\nu}}{m^{2}}\right)
\left(  \eta^{\rho\sigma}+\frac{2k^{\rho}k^{\sigma}}{m^{2}}\right)  \right)  .
\end{align*}
When $b\neq1$, the massive spin-2 field and the ghost of spin-0 are coupled.
They only decouple for $b=1$, the Fierz-Pauli choice. Fixing $b=1$ could not
be maintained at the quantum level. Canonical quantization shows that the
modes do couple at the non-linear level \cite{DamourKogan}. The tensor
$h_{\mu\nu}$ is symmetric with ten components. Unlike the massless graviton
which is protected by diffeomorphism invariance, there is no gauge symmetry
here and all component $h_{ij}$ $(i,j=1,2,3)$ propagate, giving rise to six
degrees of freedom. A massive spin-2 field has only five dynamical degrees of
freedom $\left(  2j+1=5\right)  $. This implies that there is an additional
component, a spin-0 ghost that does not decouple, except for the choice $b=1.$
The limit to the massless case $\left(  m\rightarrow0\right)  $ is singular.
This is similar to the propagator of a massive spin-1 field, which is also
singular in the $m\rightarrow0$ limit. This suggests that in order to solve
the problem with the singular zero mass limit, the mass of the spin-2 field
should be acquired through the Higgs mechanism and spontaneous symmetry
breaking. But to do this, the Lagrangian must have a gauge symmetry which
should be broken. This is not possible without extending the system in such a
way as not to increase the dynamical degrees of freedom. I\ will achieve this
by employing the following idea.

Weyl formulation of a massless graviton is based on promoting the
$SL(2,\mathbb{C})$ global invariance of the Dirac equation to a local one
\cite{Weyl} \cite{Utiyama} \cite{Kibble}%
\begin{align*}
\overline{\psi}\gamma^{\mu}\partial_{\mu}\psi &  \rightarrow\overline{\psi
}\gamma^{\mu}\nabla_{\mu}\psi,\\
\left[  \nabla_{\mu},\nabla_{\nu}\right]   &  =\frac{1}{4}R_{\mu\nu}%
^{ab}\sigma_{ab},
\end{align*}
where $\nabla_{\mu}=\partial_{\mu}+\frac{1}{4}\omega_{\mu}^{ab}\gamma_{ab}$,
and $\omega_{\mu}^{ab}$ is the spin-connection. A vierbein {$e_{\mu}^{a}$ is
introduced to insure the gauge invariance of the gravitational action, which
can be written in an index free notation}%
\[
I_{E-H}=\frac{1}{8}\int Tr\left(  \gamma_{5}e\wedge e\wedge R\right)  ,
\]
where $e=e_{\mu}^{a}\gamma_{a}dx^{\mu}$, and $R=\frac{1}{2}R_{\mu\nu}%
^{ab}\gamma_{ab}dx^{\mu}\wedge dx^{\nu}.$ The torsion defined by%
\[
T=de+\omega\wedge e+e\wedge\omega,
\]
is set to zero, which allows for $\omega_{\mu}^{ab}$ to be solved in terms of
{$e_{\mu}^{a}$ and its inverse}%
\begin{align*}
\omega_{\mu}^{ab}\left(  e\right)   &  =\frac{1}{2}e^{\nu a}e^{\rho b}\left(
\Omega_{\mu\nu\rho}\left(  e\right)  -\Omega_{\nu\rho\mu}\left(  e\right)
+\Omega_{\rho\mu\nu}\left(  e\right)  \right)  ,\\
\Omega_{\mu\nu\rho}\left(  e\right)   &  =\left(  \partial_{\mu}e_{\nu}%
^{c}-\partial_{\nu}e_{\mu}^{c}\right)  e_{\rho c}.
\end{align*}
Substituting this value of $\omega_{\mu}^{ab}\left(  e\right)  $ into the
Einstein-Hilbert action gives the familiar Ricci scalar depending only on the
metric $g_{\mu\nu}=e_{\mu}^{a}e_{\nu}^{b}\eta_{ab}$. The dependence on the
antisymmetric part of $e_{\mu}^{a}$ cancels because of the $SL(2,\mathbb{C})$
gauge invariance of the action. The six gauge parameters of $SL(2,\mathbb{C})$
can be used to eliminate the six components in the antisymmetric part of
$e_{\mu}^{a}.$ The lesson we learn from this example is that one can extend
the symmetry of the system by enlarging the number of fields. In a special
gauge there will be no trace of the symmetry. In this example, what protects
the field $g_{\mu\nu}$ of acquiring a mass is diffeomorphism invariance of the
action. It is then clear that if we have a system of two symmetric tensors,
then diffeomorphism invariance can only protect one of them from becoming massive.

{A coupled system of one massless graviton and one massive graviton, can be
formulated as a gauge theory of $SP(4)\times SP(4)$ \cite{CSS}. }The group
$SP(4)$ results from trading the diffeomorphism transformations of $e_{\mu
}^{a}$ by a translation in internal space
\begin{align*}
\delta e_{\mu}^{a} &  =\partial_{\mu}\xi^{\nu}e_{\nu}^{a}+\xi^{\nu}%
\partial_{\nu}e_{\mu}^{a}+\omega^{ab}e_{\mu b}\\
&  =\left(  \partial_{\mu}\xi^{a}+\omega_{\mu}^{ab}\left(  e\right)  \xi
_{b}\right)  +\omega^{^{\prime}ab}e_{\mu b},
\end{align*}
where the zero torsion condition is used and
\[
\zeta^{a}=\xi^{\mu}e_{\mu}^{a},\qquad\omega^{^{\prime}ab}=\omega^{ab}-\xi
^{\nu}\omega_{\nu}^{ab}\left(  e\right)  .
\]
Therefore, one can start with the gauge fields
\begin{align*}
A_{\mu} &  =ie_{\mu}^{a}\gamma_{a}+\frac{1}{4}\omega_{\mu}^{ab}\sigma_{ab},\\
A_{\mu}^{^{\prime}} &  =ie_{\mu}^{^{\prime}a}\gamma_{a}+\frac{1}{4}\omega
_{\mu}^{^{\prime}ab}\sigma_{ab},
\end{align*}
and introduce a pair of Higgs fields $G_{1}$ and $G_{2}$ transforming under
the product representation of {$SP(4)\times SP(4).$} Imposing 14 constraints
on $G_{1}$ \ and $G_{2}$ of the form
\[
Tr\left(  \left(  G_{i}\widetilde{G_{i}}\right)  ^{n}\right)  =c_{ni}%
,\hspace{0.5in}i=1,2,
\]
breaks the symmetry spontaneously {$SP(4)\times SP(4)\rightarrow
SL(2,\mathbb{C})$ through a non-linear realization \cite{CWZ}. }An action of
the form{%
\begin{align*}
&  \int Tr\left(  \alpha G_{1}\widetilde{G}_{2}F\wedge F+\alpha^{^{\prime}%
}\widetilde{G}_{1}G_{2}F^{^{\prime}}\wedge F^{^{\prime}}\right.  \\
&  \hspace{0.3in}\left.  +\beta\nabla G_{1}\wedge\nabla\widetilde{G}_{1}%
\wedge\nabla G_{1}\wedge\nabla\widetilde{G}_{2}+\beta^{^{\prime}}\nabla
G_{2}\wedge\nabla\widetilde{G}_{2}\wedge\nabla G_{2}\wedge\nabla\widetilde
{G}_{1}\right)  ,
\end{align*}
}where
\begin{align*}
F  & =dA+A\wedge A\\
\widetilde{G}  & =CG^{T}C^{-1}%
\end{align*}
$C$ \ being the charge conjugation matrix. {Analysis of the quadratic part of
this action reveals that one combination of $e_{\mu}^{a}$ and$e_{\mu
}^{^{\prime}a}$ is massless while the other combination is massive. In the
unitary gauge we can choose%
\begin{align*}
G_{1} &  =a,\\
G_{2} &  =b\gamma_{5}.
\end{align*}
To illustrate how the consistency of massive gravitons is solved, we consider
only one gauge group $SP(4)$ with gauge field $A_{\mu}$ and a Higgs multiplet
$G$}{%
\[
G=\phi i\gamma_{5}+v_{a}\gamma_{a}\gamma_{5}.
\]
This is }subject to the constraint
\[
{Tr\left(  G^{2}\right)  =-4a^{2},}%
\]
{which in component form reads }%
\[
{\phi^{2}+v_{a}v^{a}=a^{2}.}%
\]
The gauge transformations{%
\begin{align*}
\delta\phi &  =\omega^{a}v_{a},\\
\delta v_{a} &  =\omega_{a}\phi+\omega_{ab}v^{b},
\end{align*}
}allows to choose the unitary gauge where {%
\[
v_{a}=0,\hspace{0.3in}\phi=a.
\]
An invariant action for the massive spin-2 field is \cite{spontaneous}} {%
\begin{align*}
&  \int Tr\left(  GF\wedge F+G\nabla G\nabla G\nabla G\nabla G\right)  \\
&  +\int d^{4}x\sqrt{g}\left(  g^{\mu\rho}g^{\nu\sigma}-g^{\mu\nu}%
g^{\rho\sigma}\right)  H_{\mu\nu},
\end{align*}
}where
\[
{H_{\mu\nu}=Tr\left(  \nabla_{\mu}G\nabla_{\nu}G\right)  .}%
\]
In the unitary gauge this gives the Fierz-Pauli action for a symmetric tensor
$H_{\mu\nu}=h${$_{\mu\nu}=e_{\mu}^{a}e_{\nu a}$ :}%
\[
\int d^{4}x\sqrt{h}(R(h)+\Lambda)+m^{2}\int d^{4}x\sqrt{g}\left(  g^{\mu\rho
}g^{\nu\sigma}-g^{\mu\nu}g^{\rho\sigma}\right)  h_{\mu\nu}h_{\rho\sigma}.
\]
{ }It is also possible to analyze this action in the non-unitary gauge, where
the components $\phi$ and $v_{a}$ are kept. {Because of the gauge invariance
of the field $e_{\mu}^{a}$ }%
\[
{{\delta e_{\mu}^{a}=\partial_{\mu}\omega^{a}+\cdots}}%
\]
{ the dynamical degrees of $e_{\mu}^{a}$ will be identical to those of the
massless graviton, thus describing helicities $+2$ and $-2$ only.} The three
independent components of $v_{a}$ will describe the helicities $+1,$ $0,$ $-1$
{%
\[
\int d^{4}x\sqrt{g}\left(  \left(  g^{\mu\rho}g^{\nu\sigma}-g^{\mu\nu}%
g^{\rho\sigma}\right)  \partial_{\mu}v_{\nu}-m^{2}v_{a}v^{a}\right)
\]
}The other helicity $0$ in $h_{\mu\nu}$ is present in $\phi$ which couples as
a scale factor{%
\[
\int d^{4}xe\,\phi^{3}\left(  R\left(  e\right)  +\Lambda\right)
\]
}The ill behaved propagator can be avoided by working in the non-unitary
gauge, where every helicity of the 6 degrees of freedom present in the
symmetric tensor $h_{\mu\nu}$ is represented with an independent field. {The
discontinuity in the propagator is related to the strong coupling of the
scalar longitudinal component of the graviton $\phi.$ One can show that the
theory is well behaved below the cut-off scale \cite{arkani}.}

\section{Matrix Gravity}

{In D-branes, coordinates of space-time become noncommuting and $U(N)$
matrix-valued \cite{Matrix} }%
\[
\left[  X^{i},X^{j}\right]  \neq0.
\]
{ A metric on such spaces will also become matrix-valued. For example in the
case of D-0 branes a matrix model action takes the form \cite{D0}}%
\[
Tr\left(  G_{ij}\left(  X\right)  \partial_{0}X^{i}\partial_{0}X^{j}\right)  .
\]
{At very short distances coordinates of space-time can become noncommuting and
represented by matrices. One may have to use the tools of noncommutative
geometry of} {Alain Connes \cite{Connes}. }

{Developing differential geometry on such spaces is ambiguous. Defining
covariant derivatives, affine connections, contracting indices, will all
depend on the order these operations are performed because of
noncommutativity. Some of these developments lead to inconsistencies such as
the occurrence of higher spin fields \cite{maluf}. In many cases studies were
limited to abelian (commuting) matrices with Fierz-Pauli interactions
\cite{wald}.} More recently the spectral approach was taken by Avramidi
\cite{Avramidi} which implies a well defined order for geometric constructs.
{Experimentally \cite{PDG}, there is only one massless graviton. Therefore in
a consistent $U(N)$ matrix-valued gravity only one massless field should
result with all others corresponding to massive gravitons. The masses of the
gravitons should be acquired through the Higgs mechanism.}

{The lesson we learned in the last section is that one should start with a
large symmetry and break it spontaneously. The minimal non-trivial extension
of $SL(2,\mathbb{C})$ and $U(N)$ is $SL(2N,\mathbb{C}).$ This is a non-compact
group. It can be taken as a gauge group only in the first order formalism, in
analogy with $SL(2,\mathbb{C}).$}{ The vierbein $e_{\mu}^{a}$ and the
spin-connection $\omega_{\mu}^{ab}$ are conjugate variables related by the
zero torsion condition. The number of conditions in $T_{\mu\nu}^{a}=0$ is
equal to the number of independent components of $\omega_{\mu}^{ab}$, which
can be determined completely in terms of $e_{\mu}^{a}.$ }The $SL(2N,\mathbb{C}%
)$ gauge field can be expanded in the Dirac basis in the form%
\[
A_{\mu}=ia_{\mu}+\gamma_{5}b_{\mu}+\frac{i}{4}\omega_{\mu}^{ab}\sigma_{ab},
\]
where {%
\begin{align*}
a_{\mu} &  =a_{\mu}^{I}\lambda^{I},\hspace{0.3in}b_{\mu}=b_{\mu}^{I}%
\lambda^{I},\hspace{0.3in}I=1,\cdots,N^{2}-1,\\
\omega_{\mu}^{ab} &  =\omega_{\mu}^{abi}\lambda^{i},\hspace{0.3in}i=0,I.
\end{align*}
and }$\lambda^{i}$ are the $U(N)$ Gell-Mann matrices. The analogue of $e_{\mu
}^{a}\gamma_{a}$ is {%
\[
L_{\mu}=e_{\mu}^{a}\gamma_{a}+f_{\mu}^{a}\gamma_{5}\gamma_{a},
\]
}where $e_{\mu}^{a}$ and $f_{\mu}^{a}$ are $U(N)$ matrices. This is equivalent
to having complex matrix gravity. The zero torsion condition {%
\[
T=dL+LA+AL=0,
\]
}will give two sets of conditions {%
\[
T_{\mu\nu}^{a}=0,\hspace{0.3in}T_{\mu\nu}^{a5}=0,
\]
}which will overdetermined the variables $\omega_{\mu}^{ab}.$

{The correct approach \cite{colored} is to consider $SL(2N,\mathbb{C}%
)\mathbb{\times}SL(2N,\mathbb{C}),$ or equivalently the complex extension of
$SL(2N,\mathbb{C)}$ as was done by Isham, Salam and Strathdee \cite{ISS2} for
the massive spin-2 nonets. In this case{%
\[
a_{\mu}=a_{\mu}^{1}+ia_{\mu}^{2},\hspace{0.3in}b_{\mu}=b_{\mu}^{1}+ib_{\mu
}^{2},\hspace{0.3in}\omega_{\mu}^{ab}=B_{\mu}^{ab}+iC_{\mu}^{ab},
\]
}and the torsion zero constraints are enough to determine $B_{\mu}^{ab}$ and
$C_{\mu}^{ab}$ in terms of $e_{\mu}^{a}$, $f_{\mu}^{a},$ $a_{\mu}$ and
$b_{\mu}.$ One can write, almost uniquely, a metric independent gauge
invariant action which will correspond to massless $U(N)$ gravitons{%
\[
\int\limits_{M}Tr\left(  i\left(  \alpha+\beta\gamma_{5}\right)  LL^{^{\prime
}}F+i\left(  \overline{\alpha}+\overline{\beta}\gamma_{5}\right)  L^{^{\prime
}}L\overline{F}+\left(  i\lambda+\gamma_{5}\eta\right)  LL^{^{\prime}%
}LL^{^{\prime}}\right)  ,
\]
}where $L^{^{\prime}}$ is related to $L.$ For illustration, the form of this
action in the }$N=1$ \ case is
\begin{align*}
&  -\frac{1}{2}\int\limits_{M}d^{4}x\epsilon^{\mu\nu\kappa\lambda}\left(
\left(  \left(  \alpha_{2}-\beta_{1}\right)  e_{\mu a}e_{\nu b}+\frac{1}%
{2}\left(  \alpha_{1}+\beta_{2}\right)  \epsilon_{abcd}e_{\mu}^{c}e_{\nu}%
^{d}\right)  B_{\kappa\lambda}^{ab}\right.  \\
&  \qquad\qquad+\left(  \left(  \alpha_{2}+\beta_{1}\right)  f_{\mu a}f_{\nu
b}-\frac{1}{2}\left(  \alpha_{1}-\beta_{2}\right)  \epsilon_{abcd}f_{\mu}%
^{c}f_{\nu}^{d}\right)  C_{\kappa\lambda}^{ab}\\
&  \qquad\qquad\left.  +\epsilon_{abcd}\left(  \left(  \lambda-\eta\right)
e_{\mu}^{a}e_{\nu}^{b}e_{\kappa}^{c}e_{\lambda}^{d}+\left(  \lambda
+\eta\right)  f_{\mu}^{a}f_{\nu}^{b}f_{\kappa}^{c}f_{\lambda}^{d}\right)
\right)  .
\end{align*}
{To give masses to the spin-2 fields, introduce the Higgs fields $H$ and
$H^{^{\prime}}$ transforming as $L$ and $L^{^{\prime}}$ and constrained in
such a way as to break the symmetry non-linearly from $SL(2N,\mathbb{C)\times
}SL(2N,\mathbb{C)}$ to $SL(2,\mathbb{C)}.$ We can add the mass terms{%
\[
\int\limits_{M}Tr\left(  \left(  i\tau+\gamma_{5}\xi\right)  LH^{^{\prime}%
}LH^{^{\prime}}LL^{^{\prime}}+\left(  i\rho+\gamma_{5}\xi\right)
HL^{^{\prime}}HL^{^{\prime}}LL^{^{\prime}}\right)  .
\]
}Some of the relevant terms in the quadratic parts of the action are, in
component form \cite{colored},}{%
\begin{align*}
&  \int d^{4}x\epsilon^{\mu\nu\kappa\lambda}\epsilon_{abcd}Tr\left(
\alpha_{1}\left\{  E_{\mu}^{a},E_{\nu}^{^{\prime}a}\right\}  a_{\kappa\lambda
}^{2}+\alpha_{2}\left\{  F_{\mu}^{a},F_{\nu}^{^{\prime}a}\right\}
b_{\kappa\lambda}^{2}\right.  \\
&  \hspace{0.3in}\hspace{0.3in}+\beta_{1}\left\{  E_{\mu}^{a},E_{\nu
}^{^{\prime}b}\right\}  B_{\kappa\lambda}^{cd}+\beta_{2}\left\{  F_{\mu}%
^{a},F_{\nu}^{^{\prime}b}\right\}  C_{\kappa\lambda}^{cd}+\\
&  \left.  +\gamma_{1}E_{\mu}^{a}E_{\nu}^{^{\prime}b}E_{\kappa}^{c}E_{\lambda
}^{^{\prime}d}+\gamma_{2}F_{\mu}^{a}F_{\nu}^{^{\prime}b}F_{\kappa}%
^{c}F_{\lambda}^{^{\prime}d}+\delta_{1}E_{\mu}^{a}E_{\nu}^{^{\prime}%
b}E_{\kappa}^{c}F_{\lambda}^{d}+\delta_{2}F_{\mu}^{a}F_{\nu}^{^{\prime}%
b}F_{\kappa}^{c}E_{\lambda}^{d}\right)  .
\end{align*}
This action is complicated because all expressions are matrix valued.
Equations are solved perturbatively. The action can be determined to second
order in the fields, and the spectrum found to be given by two sets of $SU(N)$
matrix-valued massive gravitons, plus two singlets of gravitons, one massless
and the other is massive, as well as $SU(N)\times SU(N)$ gauge fields.}

{We decompose $E_{\mu a}^{I}$ into symmetric and antisymmetric parts}{%
\[
E_{\mu a}^{I}=S_{\mu a}^{I}+T_{\mu a}^{I},
\]
}{where $S_{\mu a}^{I}=S_{a\mu}^{I}$ is symmetric and $T_{\mu a}^{I}=-T_{a\mu
}^{I}$ is antisymmetric. The symmetric part propagates while the antisymmetric
part $T_{\mu\nu}^{I}$ couples to the Yang-Mills fields and act as auxiliary
fields to give them kinetic energies. For example besides the quadratic terms
for $T^{\mu\nu I}$ coming from the mass terms, we have} {%
\[
\int\limits_{M}d^{4}x\left(  \partial_{\mu}a_{\nu}^{1I}-\partial_{\nu}a_{\mu
}^{1I}\right)  T^{\mu\nu I},
\]
}{as well as similar couplings to $a_{\mu}^{2I},\ b_{\mu}^{1I},\ b_{\mu}^{2I}%
$. }By eliminating the field $T_{\mu\nu}^{I}$ the fields {$a_{\mu}%
^{1I},\ a_{\mu}^{2I},\ b_{\mu}^{1I},\ b_{\mu}^{2I}$} would acquire the regular
$SU(N)$ Yang-Mills gauge field strengths. {A detalied study of this system is
carried in \cite{colored}.}

\section{Conclusions}

{The novel features of this action are:}

\begin{itemize}
\item {The massless graviton is not introduced as a background metric but is a
part of the gravitons matrix.}

\item {Mass is generated spontaneously for the spin-2 fields.}

\item {The massive gravitons interact in a non-trivial way.}

\item {Correct kinetic energies for the spin-2 fields are generated, although
the gauge group is noncompact. This is achieved by utilizing the first order
formalism.}

\item {Kinetic energies for the non-abelian $SU(N)$ gauge fields are generated
by couplings to the antisymmetric parts of the matrix-vierbeins.}

\item {With the requirements of gauge invariance and the restriction that the
action is a four-form, the action is almost unique and is unambiguous.}

\item {It is possible to formulate a consistent theory of matrix gravity based
on $SL(2N,\mathbb{C})\otimes SL(2N,\mathbb{C})$ gauge symmetry spontaneously
broken to $SL(2,\mathbb{C}).$}

\item {The theory unifies the massless graviton with colored massive gravitons
with $SU(N)\times SU(N)$ symmetry as well as with gauge fields.}
\end{itemize}

{Although the gravitons are promoted to become matrix valued, the coordinates
are not. Only the diagonal component of the coordinates are kept. In a more
general treatment, the $4N^{2}$ coordinates $X^{\mu}$ should be used instead
of the 4 coordinates $x^{\mu}$ used here. It is important to learn how to
adopt this construction to the noncommutative geometry of Alain Connes based
on spectral data and to construct Dirac operators for such spaces.}

\section{Acknowledgments}

This contribution is dedicated to Pran Nath, a long time friend and
collaborator, on the occasion of his 65$^{\text{th}}$ birthday. This work is
supported in part by National Science Foundation Grant No. Phys-0314616.

\end{document}